\documentclass {article}
\usepackage {epsfig}
\newcommand {\bc}{\begin {center}}
\newcommand {\ec}{\end {center}}
\newcommand {\be}{\begin {equation}}
\newcommand {\ee}{\end {equation}}
\newcommand {\beq}{\begin {eqnarray}}
\newcommand {\eeq}{\end {eqnarray}}

\newcommand {\ovl}{\overline}
\newcommand {\sign}{\mathop {\rm sgn}\nolimits}

\def\disp{\displaystyle}
\def\prt{\partial}

\def\intl{\int\limits}
\def\cA {{\cal A}}
\def\cB {{\cal B}}
\def\cJ {{\cal J}}
\def\cH {{\cal H}}
\def\cL {{\cal L}}

\def\c {{\rm c}}
\def\d {{\rm d}}

\def\i {{\rm i}}

\def\s {{\rm s}}

\def\H {{\rm H}}
\def\J {{\rm J}}

\def\tf {\tilde {f}}
\def\tH {\tilde {\cH}}

\def\tJ {\tilde {\cJ}}
\def\as {{\rm as}}
\def\DR#1#2{\frac {\d#1}{\d#2}}
\def\Dr#1#2{\frac {\prt#1}{\prt#2}}
\begin {document}
\hoffset-3cm
\voffset-2cm

\bc
{\large{\bf TIME DEPENDENT MONOCHROMATIC SCATTERING OF RADIATION 
IN ONE DIMENSIONAL MEDIA: ANALYTICAL AND NUMERICAL SOLUTIONS}}
\ec

\bc
{\large{\bf D.I.Nagirner\footnote{e-mail: dinagirner@gmail.com} and 
S.L.Kirusheva\footnote{e-mail: skirusheva@mail.ru} }}
\ec

\bc {V.~V.~Sobolev Institute of Astronomy, St. Petersburg 
State University, Russia}
\ec
\bc
Summary
\ec

In order to choose a numerical method for solving the time dependent equations
of radiative transport, we obtain an exact solution for the time dependent
radiation field in a one dimensional infinite medium with monochromatic,
isotropic scattering for sources with an arbitrary spatial distribution and an
arbitrary time variation of their power. The Lax--Wendroff method seems to be
the most suitable. Because it is assumed that radiation delay is caused by the
finite speed of light, the following difficulty arises when the numerical
method is used: the region of variation of the variables (dimensionless
coordinate $\tau$ and time $t$) is triangular (the inequality $\tau\leq t$).
This difficulty is overcome by expanding the unknown functions in series in
terms of small values of the time and coordinate. By comparing the numerical
and exact solutions for a point source with a given time dependence for its
power and with pure scattering, the steps in the variables required to obtain a
desired accuracy are estimated. This numerical method can be used to calculate
the intensity and polarization of the radiation from sources in the early
universe during epochs close to the recombination epoch.

\bigskip


1. {\it Introduction.} Analytical and numerical solutions are given in
this paper for a simple problem of time  dependent, monochromatic scattering of
radiation in a so-called one dimensional medium. 

An idealized medium is refered to as one dimensional if a photon, either
preserves the direction of its motion upon scattering in this medium, or its
direction of motion reverses (backward-forward scattering), so that it
propagates along a straight line. The assumption of a one dimensional medium is
equivalent to assuming that the scattering takes place with an indicatrix equal
to the sum of spike (delta-function) indicatrices. This kind of scattering
problems was examined in the early development of radiative transport
theory \cite{VVS,nonstat,INM1,INM2,NBE,AGN}. The description of scattering in
an ordinary (three dimensional) medium reduces approximately to the transport
equations for a one dimensional medium \cite{VVS}.

Here we examine the propagation of radiation in a one dimensional medium with
invariant optical properties. It is assumed that the scattering is isotropic,
monochromatic and, as usually assumed for monochromatic scattering, such that
the time delay of the photons is caused by the time they spend in transit,
while scattering events are instantaneous. The sources of primary radiation can
be nonisotropic, with an arbitrary spatial distribution, and have an arbitrary
time dependence. 

Since the medium is stationary, we can introduce the optical path $\tau$,
measured from some point in units of the mean free path of the photons. We
restrict ourselves to scattering in an infinite medium, where the optical path
$\tau$ ranges from $-\infty$ to $\infty$. The time $t$ is also measured in
units of the mean time between collisions. Sources can act from time
$t=-\infty$, when there is no radiation field. The solution of this problem is
obtained here in an explicit analytical form.

The main object of attention is a point source with a given time dependence of
its radiated power. The existence of an exact solution has made it possible
to choose a numerical method for solving equations similar to those for
scattering in a one dimensional medium. It is planned to use this numerical
method in a separate article for calculating the evolution of the intensity and
polarization of radiation from sources in epochs of the early universe close to
the recombination epoch (see \cite{ZN} and \cite{{peacock}}, for example).

\medskip

2. {\it Basic equations.} We denote the intensity of the radiation
propagating in the direction of increasing and decreasing optical depths
by $I_{+}(t,\tau)$ and $I_{-}(t,\tau)$, respectively. Then the two equations
for the evolution of the radiation field can be written in the form
\be \label{eq:eqtimeIpm}
\dot {I}_\pm(t,\tau)\pm I'_\pm(t,\tau)=-I_\pm(t,\tau)+B_\pm(t,\tau).
\ee
Here a time derivative is indicated by an overhead dot and a derivative with
respect to the optical path length, by a prime. The two source functions, as
always, consist of two parts, one characterizing the power of the primary
source and the other describing the scattering:
\be \label{eq:Bpmdef}
B_\pm(t,\tau)=g_\pm(t,\tau)+\lambda\cJ_\c(t,\tau),
\ee
where $\lambda$ is the probability of survival of a photon in each scattering
event. Here we have introduced the average intensity $\cJ_\c(t,\tau)$. We can
immediately introduce the radiative flux, as well:
\be \label{eq:ovlJH}
\cJ_\c(t,\tau)=\frac {I_{+}(t,\tau)+I_{-}(t,\tau)}{2},\quad
\cH_\c(t,\tau)=\frac {I_{+}(t,\tau)-I_{-}(t,\tau)}{2}.
\ee

Adding and subtracting Eqs. (\ref{eq:eqtimeIpm}) yield equation for the
average intensity and the flux: 
\be \label{eq:eqHJ}
\dot {\cJ_\c}(t,\tau)+\cH'_\c(t,\tau,t)+(1-\lambda)\cJ_\c(\tau,t)=
f_\J(\tau,t),\quad\dot {\cH_\c}(t,\tau)+\cJ'_\c(t,\tau)+\cH_\c(t,\tau)=
f_\H(t,\tau),
\ee
where 
\be \label{eq:fJfHdef}
f_\J(t,\tau)=\frac {g_{+}(t,\tau)+g_{-}(t,\tau)}{2},\quad
f_\H(t,\tau)=\frac {g_{+}(t,\tau)-g_{-}(t,\tau)}{2}.
\ee

\medskip

3. {\it Laplace transform.} An efficient way of solving time dependent
problems is to use Laplace transforms with respect to time. We shall indicate
the transform with a tilde over the transformed function:
\be \label{eq:tJtHdef}
\tJ(\tau,s)=\intl_{-\infty}^\infty\cJ_\c(t,\tau)e^{-st}\d t,\quad
\tH(\tau,s)=\intl_{-\infty}^\infty\cH_\c(t,\tau)e^{-st}\d t.
\ee
Applying the transform to Eqs. (\ref{eq:eqHJ}) for zero initial conditions (at
$t=-\infty$) yields 
\be \label{eq:eqJHLapl}
\tH'(\tau,s)=-(s+1-\lambda)\tJ(\tau,s)+\tf_\J(\tau,s),\quad
\tJ'(\tau,s)=-(s+1)\tH(\tau,s)+\tf_\H(\tau,s).
\ee

Let us find a solution to Eqs. (\ref{eq:eqJHLapl}) assuming that the transform
parameter $s$ is real and non-negative. Finding the general solution of the
homogeneous equation and a particular solution for the non-homogeneous
equations, while noting that the solutions must be finite at
$\tau\to\pm\infty$, we obtain
\beq \label{eq:tJres}
& \strut\disp \tJ(\tau,s)=\frac {1}{2}\intl_{-\infty}^\infty
e^{-k|\tau-\tau_1|}\left[\frac {s+1}{k}\tf_\J(\tau_1,s)+\tf_\H(\tau_1,s)
\sign(\tau-\tau_1)\right]\d\tau_1, & \\ \label{eq:tHres}
& \strut\disp \tH(\tau,s)=\frac {1}{2}\intl_{-\infty}^\infty
e^{-k|\tau-\tau_1|}\left[\tf_\J(\tau_1,s)\sign(\tau-\tau_1)+
\frac {k}{s+1}\tf_\H(\tau_1,s)\right]\d\tau_1. &
\eeq
Here the positive root $k=\sqrt{(s+1)(s+1-\lambda)}$.

\medskip

4. {\it Inversion of the transforms.} In order to invert the Laplace
transforms, we use Eqs. 4.17 (5)-(9) of the Bateman and Erdelyi handbook
\cite{BE}. Excluding the singular solutions which arise on inverting the parts
of the transforms which behave as $e^{-(s+1-\lambda/2)|\tau-\tau_1|}$, we
write the result of the inversion in the form
\beq \label{eq:Jttaufgres}
& \strut\disp \cJ_\c(t,\tau)=\frac {1}{2}\intl_{-\infty}^\infty\d\tau_1
e^{-(1-\lambda/2)|\tau-\tau_1|}\left[f_\J(t-|\tau-\tau_1|,\tau_1)+
f_\H(t-|\tau-\tau_1|,\tau_1)\sign(\tau-\tau_1)\right]+ & \nonumber \\
& \strut\disp +\frac {\lambda}{4}\intl_{-\infty}^\infty\d\tau_1
\intl_{-\infty}^{t-|\tau-\tau_1|}\d t_1 e^{-(1-\lambda)(t-t_1)}\left[
G_{\J\J}(\lambda(t-t_1),\lambda(\tau-\tau_1))f_\J(t_1,\tau_1)+
G_{\J\H}(\lambda(t-t_1),\lambda(\tau-\tau_1))f_\H(t_1,\tau_1)\right],
 & \\ \label{eq:Httaufgres}
& \strut\disp \cH_\c(t,\tau)=\frac {1}{2}\intl_{-\infty}^\infty\d\tau_1
e^{-(1-\lambda/2)|\tau-\tau_1|}\left[f_\J(t-|\tau-\tau_1|,\tau_1)
\sign(\tau-\tau_1)+f_\H(t-|\tau-\tau_1|,\tau_1)\right]+ & \nonumber \\
& \strut\disp +\frac {\lambda}{4}\intl_{-\infty}^\infty\d\tau_1
\intl_{-\infty}^{t-|\tau-\tau_1|}\d t_1 e^{-(1-\lambda)(t-t_1)}\left[
G_{\H\J}(\lambda(t-t_1),\lambda(\tau-\tau_1))f_\J(t_1,\tau_1)+
G_{\H\H}(\lambda(t-t_1),\lambda(\tau-\tau_1))f_\H(t_1,\tau_1)\right]. &
\eeq
This solution is expressed in terms of the Green functions 
\beq \label{eq:GJJ}
& \strut\disp  G_{\J\J}(t,\tau)=e^{-t/2}\left[\frac {t}{w}
I_1\left(\frac {w}{2}\right)+I_0\left(\frac {w}{2}\right)\right],
 & \\ \label{eq:GHH}
& \strut\disp G_{\H\H}(t,\tau)=e^{-t/2}\left[\frac {t}{w}
I_1\left(\frac {w}{2}\right)-I_0\left(\frac {w}{2}\right)\right],
 & \\ \label{eq:GJH}
& \strut\disp G_{\J\H}(t,\tau)=G_{\H\J}(\tau,t)=e^{-t/2}\frac {\tau}{w}
I_1\left(\frac {w}{2}\right). &
\eeq
      
In these formulas $I_n(z)$ is the Bessel function of imaginary argument and
$w=\sqrt{t^2-\tau^2}$. The Green functions have  meaning for $t\geq|\tau|$,
which sets the upper limit of integration on $t_1$ in Eqs.
(\ref{eq:Jttaufgres})--(\ref{eq:Httaufgres}). The values of the functions at
the boundary of the region in which they are defined, i.e., $t=|\tau|$, are 
\beq \label{eq:ultvalues}
G_{\J\J}(|\tau|,\tau)=e^{-|\tau|/2}\left(\frac {|\tau|}{4}+1\right),\quad
G_{\H\H}(|\tau|,\tau)=e^{-|\tau|/2}\left(\frac {|\tau|}{4}-1\right),\quad
G_{\J\H}(|\tau|,\tau)=e^{-|\tau|/2}\frac {\tau}{4}.
\eeq
It can be verified by direct substitution that the functions
(\ref{eq:GJJ})--(\ref{eq:GJH}) satisfy the homogeneous equations 
\be \label{eq:eqGGG}
\dot {G}_{\J\J}+G'_{\H\J}=0,\quad \dot {G}_{\J\H}+G'_{\H\H}=0,\quad
G_{\H\H}+\dot {G}_{\H\H}+G'_{\H\J}=0,\quad
G_{\J\H}+\dot {G}_{\J\H}+G'_{\J\J}=0,
\ee
which can be used to confirm that Eqs.
(\ref{eq:Jttaufgres})--(\ref{eq:Httaufgres}) do, indeed, determine the
solutions of Eqs. (\ref{eq:eqHJ}).

Function (\ref{eq:GJJ}) was obtained previously by Minin \cite{INM2}. 

\medskip

5. {\it Point source.} Let us consider the important special case of a
time dependent, isotropic source acting at $t\geq 0$. The following functions
apply in this case:
\be \label{eq:pointsource}
f_\J(t,\tau)=\delta(\tau)\cL(t),\quad f_\H(t,\tau)=0,
\ee
where the function $\cL(t)$ characterizes the variation of the source power 
with time. We shall assume that it is continuous and falls off rapidly
enough so that the effective duration $t_\s$ of the source is finite. The
corresponding average intensity and flux are single integrals. For
$t\geq|\tau|$,
\beq \label{eq:Jpointsource}
& \strut\disp \cJ_\c(t,\tau)=\frac {1}{2}e^{-(1-\lambda/2)|\tau|}\cL(t-|\tau|)+
J_\i(t,\tau), & \\ \label{eq:Hpointsource}
& \strut\disp \cH_\c(t,\tau)=\frac {1}{2}e^{-(1-\lambda/2)|\tau|}\cL(t-|\tau|)
\sign(\tau)+H_\i(t,\tau), &
\eeq
where the integral terms are given by 
\be \label{eq:JiHi}
J_\i(t,\tau)=\frac {\lambda}{4}\intl_0^{t-|\tau|}\d t_1 e^{-(1-\lambda)(t-t_1)}
G_{\J\J}(\lambda(t-t_1),\lambda\tau)\cL(t_1),\,\,
H_\i(t,\tau)=\frac {\lambda}{4}\intl_0^{t-|\tau|}\d t_1 e^{-(1-\lambda)(t-t_1)}
G_{\H\J}(\lambda(t-t_1),\lambda\tau)\cL(t_1).
\ee
The average intensity is continuous for all $\tau$ and $t\geq|\tau|$. For
$\tau=0$ and $\cL(t)>0$, i.e., at the site of the point source and during the
time it acts, the flux has a discontinuity: $\cH(+0,t)-\cH(-0,t)=\cL(t)$.
At the points $\tau=\pm t$, where the medium only begins to radiate, the
solutions have the following limiting values:
\be \label{eq:bondcond}
\cJ(|\tau|,\tau)=\frac {1}{2}e^{-(1-\lambda/2)|\tau|}\cL(0),\quad
\cH(|\tau|,\tau)=\cJ(|\tau|,\tau)\sign(\tau).
\ee

It is easy to determine how the solutions behave at late times compared to the
time the source is active and at large distances from it, i.e., for $t\gg t_\s$
and $t\gg\tau$. We shall assume that the ratios $t_\s/t$ and $\tau/t$ are
small, while the ratio $t_0=\tau^2/t$ is of the order of unity (i.e., $t_0/t$
is small). For $\lambda=1$, we have 
\beq \label{eq:Jiexp2}
& \strut\disp J_\i(t,\tau)\sim\frac {e^{-t_0/4}}{2\sqrt {\pi t}}\left\{\cL_0-
\left[\cL_1\left(\frac {t_0}{4}-\frac {1}{2}\right)+\cL_0\left(
\frac {t_0^2}{16}-\frac {t_0}{2}+\frac {1}{4}\right)\right]\frac {1}{t}+
\left[\cL_2\left(\frac {t_0^2}{32}-\frac {3}{8}t_0+\frac {3}{8}\right)+
 \right.\right. & \nonumber \\
& \strut\disp \left.\left. +\cL_1\left(\frac {t_0^3}{64}-\frac {11}{32}t_0^2+
\frac {21}{16}t_0-\frac {3}{8}\right)+\cL_0\left(\frac {t_0^4}{512}-
\frac {t_0^3}{16}+\frac {27}{64}t_0^2-\frac {3}{8}t_0-\frac {3}{32}\right)
\right]\frac {1}{t^2}\right\}, &
\eeq
\beq \label{eq:Hiexp2}
& \strut\disp H_\i(t,\tau)\sim\frac {e^{-t_0/4}}{4\sqrt {\pi t}}\frac {\tau}{t}
\left\{\cL_0-
\left[\cL_1\left(\frac {t_0}{4}-\frac {3}{2}\right)+\cL_0\left(
\frac {t_0^2}{16}-\frac {3}{4}t_0+\frac {3}{4}\right)\right]\frac {1}{t}+
\left[\cL_2\left(\frac {t_0^2}{32}-\frac {5}{8}t_0+\frac {15}{8}\right)+
 \right.\right. & \nonumber \\
& \strut\disp \left.\left. +\cL_1\left(\frac {t_0^3}{64}-\frac {15}{32}t_0^2+
\frac {45}{16}t_0-\frac {15}{8}\right)+\cL_0\left(\frac {t_0^4}{512}-
\frac {5}{64}t_0^3+\frac {45}{64}t_0^2-\frac {15}{16}t_0-\frac {15}{32}\right)
\right]\frac {1}{t^2}\right\}. &
\eeq
In these equations the moments of the source power are 
\be \label{eq:cLn}
\cL_n=\intl_0^\infty\cL(t)t^n\d t,
\ee
which are assumed to be finite. 

The main terms in the asymptotes are solutions of the diffusion equation, and a
diffusive relation holds between the average intensity and flux, i.e.,
$H(t,\tau)=-J'(t,\tau)$. The flux falls off with time substantially more
rapidly than the average intensity.

\medskip

6. {\it Direct and scattered radiation.} The solutions of Eq.
(\ref{eq:eqHJ}) for $\lambda=0$ represent the radiation propagating directly
from the source, without undergoing any scattering. For a point source, these
solutions are
\be \label{eq:directrad}
J_*(t,\tau)=\frac {1}{2}e^{-|\tau|}\cL(t-|\tau|),\quad
H_*(t,\tau)=J_*(t,\tau)\sign(\tau).
\ee
The scattered (diffuse) radiation is characterized by quantities that go to
zero for $\lambda=0$ and are equal to the differences of the corresponding
functions (\ref{eq:Jpointsource})--(\ref{eq:Hpointsource}) and
(\ref{eq:directrad}), i.e.,
\be \label{eq:scatrad}
\cJ_\d(t,\tau)=J_\s(t,\tau)+J_\i(t,\tau),\quad
\cH_\d(t,\tau)=J_\s(t,\tau)\sign(\tau)+H_\i(\tau,t),
\ee
where the expression outside the integrals is 
\be \label{eq:Jsttaudef}
J_\s(t,\tau)=\cA(\tau)\cL(t-|\tau|),\quad \cA(\tau)=
\frac {1}{2}\left(e^{-(1-\lambda/2)|\tau|}-e^{-|\tau|}\right).
\ee
The functions $\cJ_\d(t,\tau)$ and $\cH_\d(t,\tau)$ obey the equations
\be \label{eq:eqJH}
\dot {\cJ_\d}(t,\tau)+\cH'_\d(t,\tau)+(1-\lambda)\cJ_\d(t,\tau)=J_*(t,\tau),
\quad\dot {\cH_\d}(t,\tau)+\cJ'_\d(t,\tau)+\cH_\d(t,\tau)=0.
\ee
The boundary conditions for these functions are analogous to Eq.
(\ref{eq:bondcond}), but instead a single exponent it is necessary to take the
difference, i.e., to replace the factor in  $\cL(0)$ by the function
$\cA(\tau)$. At the site of the source (for $\tau=0$) the average intensity and
flux of the scattered radiation are continuous, with
\be
J_\s(t,0)=\cH_\d(t,0)=0.
\ee
The asymptotes of the functions $\cJ_\d(t,\tau)$ and $\cH_\d(t,\tau)$ coincide
with the asymptotes of the functions $J_\i(t,\tau)$ and $H_\i(t,\tau)$, since
the expressions outside the integral fall off more rapidly than the integrals.

Because of the obvious symmetry of the average intensity and the asymmetry of
the flux with respect to $\tau$, in the following we consider $\tau\geq 0$,
i.e., we consider the radiation field from one side of the source.

\medskip

7. {\it Power-law sources.} We now consider a rather general kind of
source whose power can be expanded in a power series in time near the onset of
their activity:
\be \label{eq:expcLtmu}
\cL(t)=t^\mu L(t),\quad L(t)=\sum_{l=0}^\infty L_l t^l,
\ee
where $\mu\geq 0$, while $L_0\neq 0$. For source of this type it is appropriate
to introduce the new unknown functions (with $\mu$ and $\lambda$ as parameters)
\be \label{eq:JHJmuHmu}
\cJ_\d(t,\tau)=y^\mu J(t,\tau),\quad
\cH_\d(t,\tau)=y^\mu H(t,\tau),\quad y=t-\tau.
\ee
These functions, again, obey the equations 
\be \label{eq:eqsJmuHmu}
\dot {J}+H'+\frac {\mu}{y}(J-H)+(1-\lambda)J=\frac {1}{2}e^{-\tau}L(y),\quad
\dot {H}+J'+\frac {\mu}{y}(H-J)+H=0.
\ee
The boundary conditions for these functions are derived from Eq.
(\ref{eq:bondcond}):
\be \label{eq:condbond}
J(\tau,\tau)=H(\tau,\tau)=\cA(\tau)L_0.
\ee
The integral terms near the boundary $t=\tau$ for power-law sources behave as
($y\ll 1$)
\beq \label{eq:Jitauytau}
& \strut\disp J_\i(\tau+y,\tau)\sim\frac {y}{4}\left[
\frac {L_0G_{\J\J}(\tau,\tau)}{\mu+1}+\frac {y}{\mu+2}\left(G_{\J\J}(\tau,\tau)
L_1+\frac {L_0G_{\J\J}^{(1)}(\tau)}{\mu+1}\right)\right],
 & \\ \label{eq:Hitauytau}
& \strut\disp H_\i(\tau+y,\tau)\sim\frac {y}{4}\left[
\frac {L_0G_{\H\J}(\tau,\tau)}{\mu+1}+\frac {y}{\mu+2}\left(G_{\H\J}(\tau,\tau)
L_1+\frac {L_0G_{\H\J}^{(1)}(\tau)}{\mu+1}\right)\right]. &
\eeq
Here the values of the Green functions at the boundary are given by Eq.
(\ref{eq:ultvalues}), while 
\be \label{eq:GJJtauGHJtau}
G_{\J\J}^{(1)}(\tau)=-\frac {1}{4}e^{-\tau/2}\left(1-\frac {\tau^2}{16}
\right),\quad G_{\H\J}^{(1)}(\tau)=-\frac {\tau}{8}e^{-\tau/2}\left(1-
\frac {\tau}{16}\right).
\ee

\medskip

8. {\it  Expansions for small} $t$ {\it and} $\tau$. The distinctive
feature of the problem is that the region of variation of its parameters has
the shape of an infinite triangle with boundaries in the $(t,\tau)$ plane
formed by the rays $\tau=0$ and $\tau=t$  emerging from the coordinate origin.
Near the vertex of this triangle it is extremely difficult to discretize the
variables, so that it is impossible (or very difficult) to use a numerical
method from the time of the very onset of the source, $t=0$. Thus, it makes
sense to move slightly away from the vertex point to some other time. Near this
point the variables have small values and, if the function  $L(t)$ has several
derivatives at $t=0$, then all the functions, including the unknown functions,
can be expanded in Taylor series in the neighborhood of this vertex. If $L(t)$
is analytic, then expansions in power series are possible. Note that our
functions are not analytic with respect to the argument $\tau$, since they
depend on $|\tau|$. Thus, the power law expansion, we shall obtain for
$\tau\geq 0$, cannot be extended to negative $\tau$.

It is easy to see that for a power law source whose power is given by Eqs.
(\ref{eq:expcLtmu}),  because the equations are linear, it is possible to
resolve components of the radiation field belonging to different powers in the
expansion of the function $L$ and proportional to the coefficients $L_l$, i.e.,
one can write the average intensity and flux in the form 
\be \label{eq:JHJlHl}
J(t,\tau)=\sum_{l=0}^\infty y^l L_l J_l(\tau,y),\quad
H(t,\tau)=\sum_{l=0}^\infty y^l L_l H_l(\tau,y),
\ee
where  $y=t-\tau$. 

Substituting Eqs. (\ref{eq:JHJlHl}) in Eqs. (\ref{eq:eqsJmuHmu}), we find
equations for the functions which are the coefficients of the expansions: 
\be \label{eq:eqsJlHlttau}
\dot {J}_l+H'_l+(1-\lambda)J_l+\frac {\mu+l}{y}(J_l-H_l)=\frac {\lambda}{2}
e^{-\tau},\quad\dot {H}_l+J'_l+H_l-\frac {\mu+l}{y}(J_l-H_l)=0.
\ee
We now transform to the variables $\tau$ and $y$, on which these functions
depend. Then the new equations take the form
\be \label{eq:eqsJlHltauy}
\Dr {H_l}{\tau}+\Dr {}{y}(J_l-H_l)+\frac {\mu+l}{y}(J_l-H_l)+(1-\lambda)J_l=
\frac {\lambda}{2}e^{-\tau},\quad
\Dr {J_l}{\tau}-\Dr {}{y}(J_l-H_l)-\frac {\mu+l}{y}(J_l-H_l)+H_l=0.
\ee
Adding these equations yields 
\be \label{eq:coroll}
\Dr {}{\tau}(J_l+H_l)+(1-\lambda)J_l+H_l=\frac {\lambda}{2}e^{-\tau}.
\ee
     
As $y\to 0$, Eqs. (\ref{eq:eqsJlHltauy}) imply that $J_l(\tau,0)=H_l(\tau,0)$,
while in the expansions (\ref{eq:JHJlHl}) only the zeroth order terms
$J_0(\tau,0)=H_0(\tau,0)$ remain. Taking the same limit in Eq.
(\ref{eq:coroll}) and using the equality above yields the following, easily
integrable linear differential equation:
\be \label{eq:DRJtau0tau}
\DR {J_0(\tau,0)}{\tau}+\left(1-\frac {\lambda}{2}\right)J_0(\tau,0)=
\frac {\lambda}{4}e^{-\tau}.
\ee
Solution of Eq. (\ref{eq:DRJtau0tau}) gives the function that depends on $\tau$
in the term under the integral in Eq. (\ref{eq:Jsttaudef}) subject to the boundary
condition (\ref{eq:condbond}): $J_0(\tau,0)=\cA(\tau)$.

It is clear from Eqs. (\ref{eq:eqsJlHlttau}) or (\ref{eq:eqsJlHltauy}) that the
index $\mu$ and number  $l$ appear in them only as a sum. Thus it is
appropriate to use the notation $\mu+l+1=z$ and regard this sum as an implicit
argument of the functions $J_l(\tau,y)$ and $H_l(\tau,y)$. Hence, we can write
the expansion for these functions at early times and for small values of the
coordinate in the form
\be \label{eq:JlHlexpan}
J_l(\tau,y)=\sum_{m=0}^\infty\sum_{j=0}^m J_{m,j}(z,\lambda)\tau^{m-j}y^j,\quad
H_l(\tau,y)=\sum_{m=0}^\infty\sum_{j=0}^m H_{m,j}(z,\lambda)\tau^{m-j}y^j,
\ee
where the parameters $\lambda$ and $z$ are indicated as arguments of the
coefficients with their two indices. In the following we shall often omit
these arguments.

Equality of these functions at $y=0$ requires that
$J_{m,0}(z,\lambda)=H_{m,0}(z,\lambda)$. The fact that the flux is zero at
$\tau=0$ leads to the following condition
\be \label{eq:Hmm0}
H_l(0,y)=\sum_{m=0}^\infty H_{m,m}(z,\lambda)y^m=0,
\ee
so that $H_{m,m}(z,\lambda)=0$ for all $m$. 

Substituting expansion (\ref{eq:JlHlexpan}) into Eq. (\ref{eq:eqsJlHltauy}), 
we obtain a pair of recurrence relations (omitting the arguments $z$ and
$\lambda$),
\beq  \label{eq:recur1}
& \strut\disp (m+1-j)H_{m+1,j}+(z+j)(J_{m+1,j+1}-H_{m+1,j+1})+
(1-\lambda)J_{m,j}=\frac {\lambda}{2}\frac {(-1)^m}{m!}\delta_{j,0},
 & \\ \label{eq:recur2}
& \strut\disp (m+1-j)J_{m+1,j}-(z+j)(J_{m+1,j+1}-H_{m+1,j+1})+H_{m,j}=0. &
\eeq

Adding these two equations yields a simple formula which follows from Eq.
(\ref{eq:DRJtau0tau}):
\be \label{eq:Jmp1j}
(m+1-j)(J_{m+1,j}+H_{m+1,j})+(1-\lambda)J_{m,j}+H_{m,j}=\frac {\lambda}{2}
\frac {(-1)^m}{m!}\delta_{j,0}.
\ee
The value $j=0$ gives
\be \label{eq:Jmp1j0}
(m+1)J_{m+1,0}+\left(1-\frac {\lambda}{2}\right)J_{m,0}=\frac {\lambda}{2}
\frac {(-1)^m}{m!},
\ee
from which we find
\be \label{eq:Jm0sol}
J_{m,0}=H_{m,0}=\frac {1}{2}\frac {(-1)^m}{m!}\left[\left(1-\frac {\lambda}{2}
\right)^m-1\right],
\ee
which is the coefficient in the expansion of $\cA(\tau)$.

Setting $m=0$ and $j=0$ in Eqs. (\ref{eq:recur1})--(\ref{eq:recur2}) gives
\be \label{eq:m0j0}
H_{1,0}+z(J_{1,1}-H_{1,1})+(1-\lambda)J_{0,0}=\frac {\lambda}{2},\quad
J_{1,0}-z(J_{1,1}-H_{1,1})+H_{0,0}=0.
\ee
Given that $J_{0,0}=H_{0,0}=0$, $\disp J_{1,0}=H_{1,0}=\frac {\lambda}{4}$,
$H_{1,1}=0$, this leads to the value of yet another coefficient:
$\disp J_{1,1}=\frac {\lambda}{4z}$.

In general, for a given value of $m$, $2(m+2)$ new coefficients
$J_{m+1,j}$ and $H_{m+1,j}$ appear, where $j=0,1,...,m+1$, but only to $2(m+1)$
equations. However, the values of $J_{m+1,0}=H_{m+1,0}$ are known, while
$H_{m+1,m+1}=0$, so that one of the equations is redundant. This equation, the
sum of a pair of equations (for $j=0$), i.e., Eq. (\ref{eq:Jmp1j0}), has been
solved for all $m$. The procedure for solving the equations for all $m$ is
the same. First we find the difference in the coefficients with one value of
$j$, and then their sum, going from smaller to larger $j$. Here $z$ enters as a
parameter.

\medskip

9. {\it Expansions of the exact solutions.} The method of obtaining
recurrence relations for the coefficients of the expansion can be applied to
more complicated and more general equations, including those with variable
coefficients in front of derivatives. We suppose to use it for calculating the
radiation fields in an expanding universe. In this work, it is possible to
obtain expansions for the functions from their explicit expressions.

Successively expanding the Bessel functions and exponents under the integral
sign in Eqs. (\ref{eq:JiHi}) and then integrating with respect to $t_1$ and
equating all the terms in the form of functions of $\tau$ and $y$, for the
average intensity we find the expansion coefficients for $m\geq 1$ and
$1\leq j\leq m$ to be
\be \label{eq:Jmjres}
J_{m,j}(z,\lambda)=\frac {\lambda}{4}\cB_j(z)\sum_{k=0}^{[(m-j)/2]}
\ovl {J}_{m-1,k}(\lambda)\frac {(m-2k-1)!}{(m-2k-j)!},
\ee
where the square brackets in the upper limit of the sum denote taking the
integer part, while 
\be \label{eq:cBjzdef}
\cB_j(z)=\sum_{i=0}^{j-1}\frac {(-1)^i}{i!(j-i-1)!(z+i)},\quad z=\mu+l+1.
\ee
The coefficient under the summation sign, which depends on $\lambda$, is given
by
\be \label{eq:Jkm}
\ovl {J}_{m,k}(\lambda)=\frac {(-1)^{k+m}}{k!}\sum_{n=k}^{[m/2]}
\frac {\lambda^{2n}\lambda_1^{m-2n}}{2^{4n}n!(n-k)!(m-2n)!}
\left(1-\frac {\lambda}{4\lambda_1}\frac {m-2n}{n+1}\right),
\ee
where $\lambda_1=1-\lambda/2$.

The coefficients in the expansion for the flux are obtained in a similar way
($m\geq 1,\,1\leq j\leq m$):
\beq \label{eq:Hmjres}
& \strut\disp H_{m+1,j}(z,\lambda)=\frac {\lambda^2}{16}\cB_j(z)
\sum_{k=0}^{[(m-j)/2]}\ovl {H}_{m-1,k}(\lambda)\frac {(m-2k-1)!}{(m-2k-j)!},
 & \\ \label{eq:Hkm}
& \strut\disp \ovl {H}_{m,k}(\lambda)=\frac {(-1)^{k+m}}{k!}\sum_{n=k}^{[m/2]}
\frac {\lambda^{2n}\lambda_1^{m-2n}}{2^{4n}(n+1)!(n-k)!(m-2n)!}. &
\eeq
The coefficients obtained from the exact formulas and from the recurrence
relations (\ref{eq:recur1})--(\ref{eq:recur2}) are the same. 

\medskip

10. {\it  Bell-shaped sources.} We now consider one concrete examples of
a time dependence for the source power: sources acting for a finite time
$t_\s$, delivering a total energy equal to 1, rising and turning off smoothly.
For a source of this type, we can assume a time dependent profile of the form
\be \label{eq:gtcont}
L(t)=c_\mu\frac {t_\s^{1-\mu}}{t^2}\left[1-\cos\left(2\pi\frac {t}{t_\s}\right)
\right]=2c_\mu\frac {t_\s^{1-\mu}}{t^2}
\sin^2\left(\pi\frac {t}{t_\s}\right),\quad 0\leq t\leq t_\s,\,\,\mu\geq 0,
\ee 
where $c_\mu$ is a normalizing coefficient.

\bc

Table 1. Estimated Accuracy of the Asymptotes at $t=20$

\smallskip

\begin {tabular*}{10.1cm}{@{\extracolsep{\fill}}|r|r|r|r|r|}
\hline
$\tau$ & $\cJ(t,\tau)$ & $J_\as(t,\tau)$ & $\cH(t,\tau)$ & $H_\as(t,\tau)$ \\
\hline
 0 & 0.062695 & 0.062687 & 0.0000000 & 0.0000000 \\
 5 & 0.046926 & 0.046928 & 0.0058817 & 0.0058821 \\
10 & 0.018610 & 0.018589 & 0.0049194 & 0.0049089 \\ 
14 & 0.048386 & 0.048526 & 0.0019490 & 0.0019418 \\
18 & 0.004726 & 0.005010 & 0.0002925 & 0.0003137 \\
\hline
\end {tabular*}

\ec

\medskip

Because the source acts for a finite time, the upper limit in the integrals of
Eqs. (\ref{eq:JiHi}) should be taken to be $\min(t-|\tau|,t_\s)$. The rest of
the equations are valid without changes. In particular, the expansions for a
bell-shaped source can be obtained by noting that in the expansion of Eq.
(\ref{eq:gtcont}), only even coefficients are nonzero:
\be \label{eq:L2lL2lp1}
L_{2l}=2L_0\frac {(-1)^l}{(2l+2)!}\left(\frac {2\pi}{t_\s}\right)^{2l},\quad
L_{2l+1}=0,\quad L_0=L(0)=2\pi^2\frac {c_\mu}{t_\s^{1+\mu}}.
\ee

\begin {figure}[h]

\centerline {\psfig {file=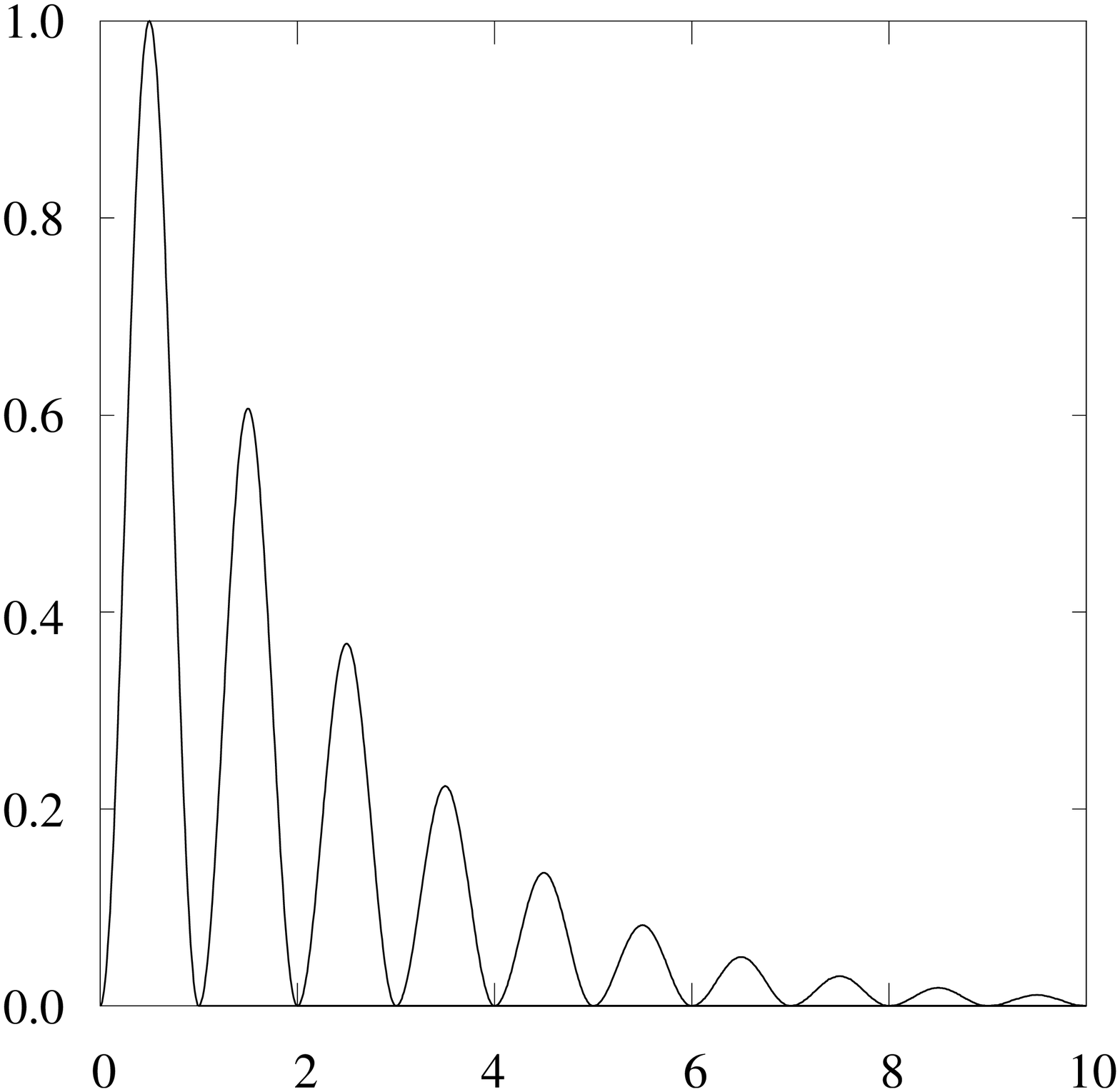,height=7cm}}

\vspace {-4mm} \hspace {130mm} $t$

\centerline {Fig. 1. The term outside the
integrals in Eqs. (\ref{eq:Jpointsource})--(\ref{eq:Hpointsource})}

\hspace {50mm} {as a function of $t$ for $t_\s=1$, $\mu=2$, $\lambda=1$ and
$\tau=0(1)9$.}
\end {figure}

%\bigskip

\begin {figure}
\centerline {\psfig {file=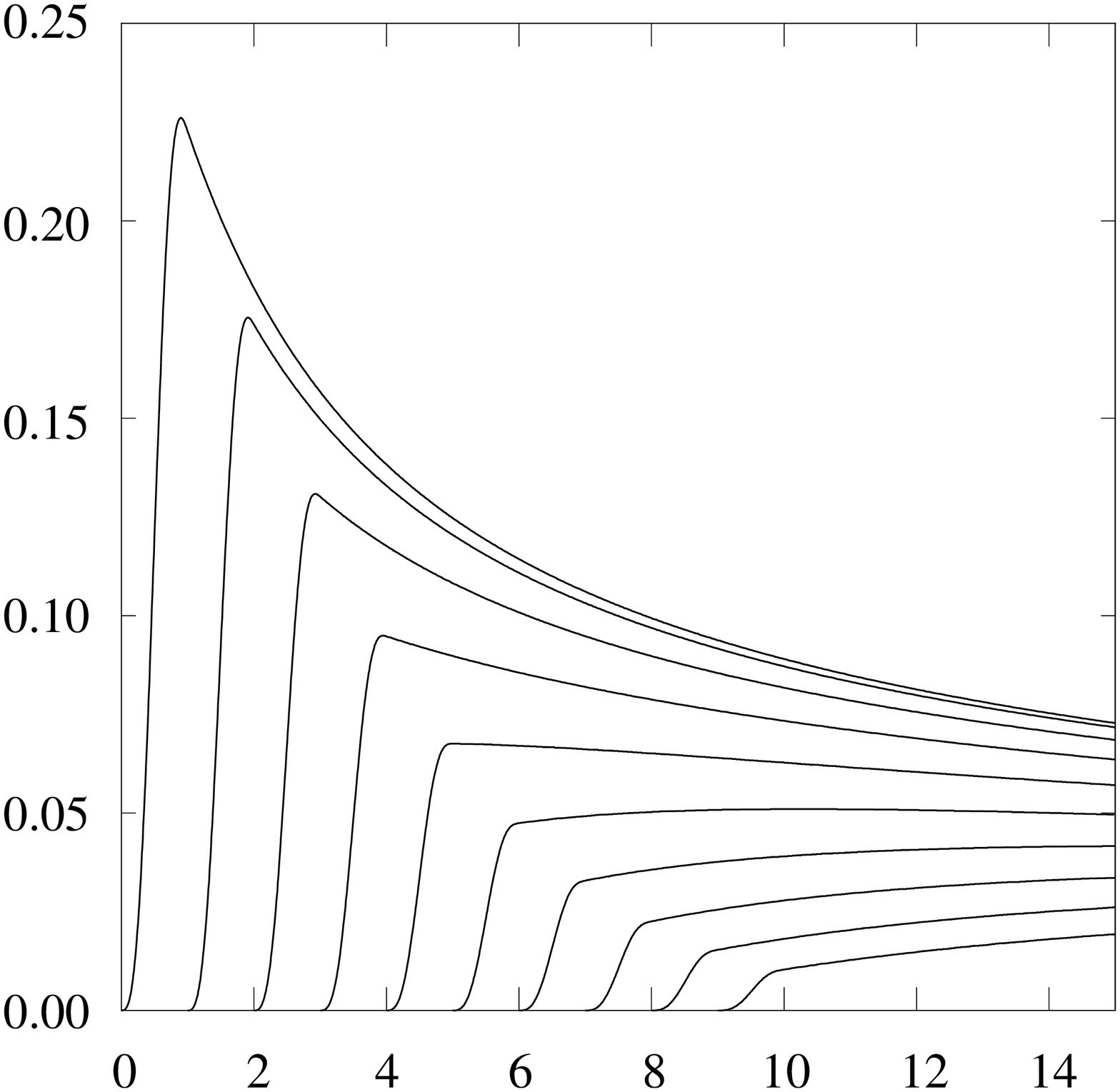,height=7cm}}

\vspace {-4mm} \hspace {130mm} $t$

\centerline {Fig.2. The integral term in $J_\i(t,\tau)$
of Eq. (\ref{eq:JiHi}) as a function of $t$}

\hspace {50mm} {for $t_\s=1$, $\mu=2$, $\lambda=1$ and
$\tau=0(1)9$.}

\bigskip

\centerline {\psfig {file=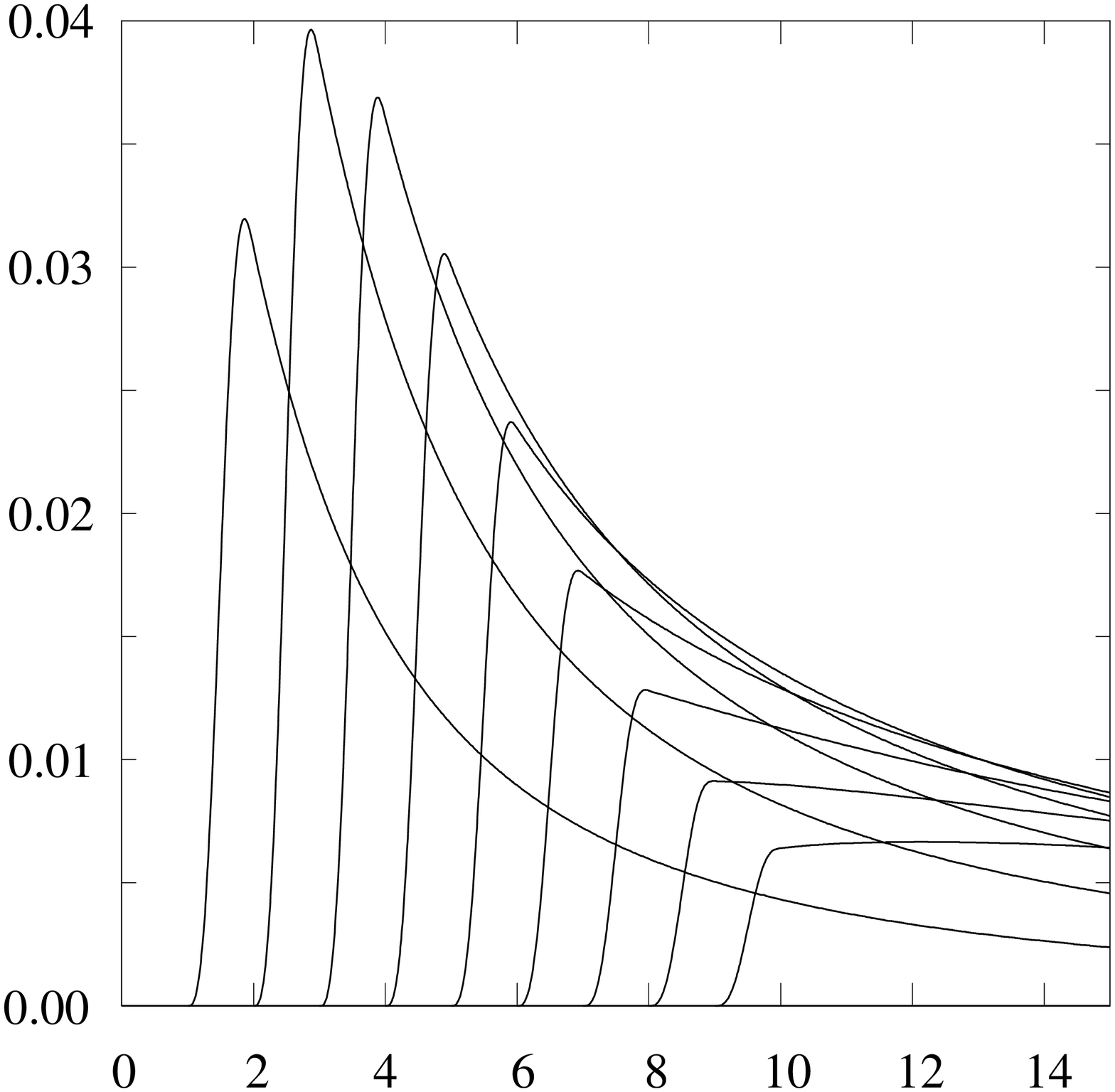,height=7cm}}

\vspace {-4mm} \hspace {130mm} $t$

\centerline {Fig. 3. The integral term in $H_\i(t,\tau)$
of Eq. (\ref{eq:JiHi}) as a function}

\hspace {50mm} {of $t$ for $t_\s=1$, $\mu=2$, $\lambda=1$ and
$\tau=0(1)9$.}

\end {figure}

The moments of the power are given by the series 
\be \label{eq:cLncmu}
\cL_n=4\pi^2c_\mu\sum_{l=0}^\infty\frac {(-1)^l}{(2l+2)!}
\frac {(2\pi)^{2l}}{2l+1+\mu+n}.
\ee
The normalizing coefficient is determined by the condition $\cL_0\!=\!1$.
In particular, $c_0\!=\!0.11223,\,c_1\!=\!0.41023,\,c_2=1$ and $c_3=2$.

As a comparison, Table 1 lists the asymptotic
(Eqs. (\ref{eq:Jiexp2})--(\ref{eq:Hiexp2})) and numerically exact values of
these quantities for  $t=20$ and a number of values of $\tau$. Naturally, the
accuracy deteriorates with increasing $\tau$. 

Figure 1 contains 10 plots of the part of the average intensity and flux
outside the integrals in Eqs. (\ref{eq:scatrad})
(function (\ref{eq:Jsttaudef})) for $\tau\geq 0$ while figures 2 and 3 show
the integrals (\ref{eq:JiHi}) for $\lambda=1$, $\mu=2$ and $t_\s=1$ as
functions of time for several values of $\tau$. The curves correspond to those
values of $\tau$ from which they begin on the abscissa. In the figures we can
see only the tendency of approaching to the asymptotes.

\medskip

11. {\it  Numerical method.} Several direct methods of solving the system
of Eqs. (\ref{eq:eqJH}) based on discretizing the equations were tried. The
most suitable was the Lax-Wendroff method. This method is usually prescribed
for partial differential equations with zero right hand sides
\cite{Num1}--\cite{Num4}, but it is easy to generalize it to systems with
nonzero right hand sides.

This method turned out to be sufficiently stable if it was applied to equations
of the form (\ref{eq:eqJH}) for a source with power given by Eq.
(\ref{eq:expcLtmu}) with integer values of the parameter $\mu$. Here we
illustrate this numerical method for the case of pure scattering, $\lambda=1$,
and $\mu=2$. For brevity we omit the index d. The equations take the form
\be \label{eq:eqscJmucHmu}
\dot {\cJ}+\cH'=\frac {1}{2}e^{-\tau}\cL(y),\quad
\dot {\cH}+\cJ'+\cH=0.
\ee

If the unknown functions have already been found at some time  $t$ for a set
of values of the coordinate  $\tau$, then according to this method the values
for subsequent times and the same coordinates are found using a second order
Taylor expansion,
\be \label{eq:JHtDeltat}
\cJ(t+\Delta t,\tau)=\cJ(t,\tau)+\dot {\cJ}(t,\tau)\Delta t+\ddot {\cJ}(t,\tau)
\frac {(\Delta t)^2}{2},\quad \cH(t+\Delta t,\tau)=\cH(t,\tau)+
\dot {\cH}(t,\tau)\Delta t+\ddot {\cH}(t,\tau)\frac {(\Delta t)^2}{2}.
\ee
The time derivatives can be expressed in terms of derivatives with respect to
the coordinate using Eqs. (\ref{eq:eqscJmucHmu}).

Let us discretize the variables, taking equal step $h$ in time and in the
coordinate: $t_i=hi$, $\tau_j=hj$, $i=0(1)i_0$, $j=0(1)i$. The region of the
discretized points is shown schematically in Fig. 4. As the initial time we
take $t_*=t_{i_*}=i_*h$, where the number $i_*$ is defined below. We calculate
the values of the unknown functions at this time using their expansions
(\ref{eq:JHJlHl}) and (\ref{eq:JlHlexpan}). The coefficients in the expansions
are found either using the recurrence relations
(\ref{eq:recur1})--(\ref{eq:recur2}) or from their exact expressions
(\ref{eq:Jmjres})--(\ref{eq:Hkm}). 

\begin {figure}[t]
\begin {picture}(150,160)(-120,0)
\put(0,0){\vector(0,1){150}}
\put(0,0){\vector(1,0){100}}
\put(-15,60){\small $\tau_{i_*}$}
\put(1,60){\line(1,1){90}}
\put(1,60){\circle*{3}}
\put(1,45){\circle*{3}}
\put(1,30){\circle*{3}}
\put(1,15){\circle*{3}}
\put(1,1){\circle*{3}}
\put(16,0){\circle*{3}}
\put(31,0){\circle*{3}}
\put(46,0){\circle*{3}}
\put(61,0){\circle*{3}}
\put(15,15){.}
\put(15,30){.}
\put(15,45){.}
\put(14,55){*}
\put(15,75){\circle*{3}}
\put(-15,90){\small $\tau_i$}
\put(15,90){.}
\put(28,-10){\small $t_{i}$}
\put(-20,-10){\small $t_{i_*}\!\!=\!hi_*$}
\put(30,90){\circle*{3}}
\put(45,105){\circle*{3}}
\put(60,120){\circle*{3}}
\put(30,15){.}
\put(30,30){.}
\put(30,45){.}
\put(30,60){.}
\put(29,70){*}
\put(43,-10){\small $t_{i+\!1}$}
\put(45,15){.}
\put(45,30){.}
\put(45,45){.}
\put(45,60){.}
\put(45,75){.}
\put(44,85){*}
\put(60,15){.}
\put(60,30){.}
\put(60,45){.}
\put(60,60){.}
\put(60,75){.}
\put(60,90){.}
\put(59,100){*}
\put(15,120){.}
\put(30,120){.}
\put(45,120){.}
\put(60,120){.}
\put(90,-15){\small $t$} \put(-15,140){\small $\tau$}
\put(90,140){\small $\tau\!=\!t$}
\end {picture}

\bigskip

\hspace {20mm} Fig. 4. The scheme for discretizing the equations.
 
\end {figure}

We denote the values of the unknown functions at the nodal points by 
$\cJ_{i,j}=\cJ(t_i,\tau_j)$ and $\cH_{i,j}=\cH(t_i,\tau_j)$ and that of the
source power, by $\disp\cL_{i-j}=\cL(h(i-j))$. The corresponding values of the
derivatives are indicated, as before, by a dot and a prime. A prime also
indicates a derivative with respect to the argument of the function $\cL(t)$.
The numerical values of the derivatives with respect to the coordinate are
calculated using the formulas
\be \label{eq:approxJH}
\cJ'_{i,j}\approx\frac {\cJ_{i,j+1}\!-\!\cJ_{i,j-1}}{2h},\,\,
\cJ''_{i,j}\approx\frac {\cJ_{i,j+1}\!-\!2\cJ_{i,j}\!+\!\cJ_{i,j-1}}{h^2},\,\,
\cH'_{i,j}\approx\frac {\cH_{i,j+1}\!-\!\cH_{i,j-1}}{2h},\,\,
\cH''_{i,j}\approx\frac {\cH_{i,j+1}\!-\!2\cH_{i,j}\!+\!\cH_{i,j-1}}{h^2}.
\ee
As a result of discretization, we obtain a system of discrete recurrence
relations of the type (for $i>i_*$,  with $j=1,...,i-1$)
\beq \label{eq:Jip1}
& \strut\disp \cJ_{i+1,j}=\frac {h}{4}e^{-jh}\left[2\cL_{i-j}+h\cL'_{i-j}
\right]-\left(1-\frac {h}{2}\right)\frac {\cH_{i,j+1}-\cH_{i,j-1}}{2}+
\frac {\cJ_{i,j+1}+\cJ_{i,j-1}}{2}, & \\ \label{eq:Hip1}
& \strut\disp \cH_{i+1,j}=\frac {h^2}{4}e^{-jh}\left(\cL_{i-j}+\cL'_{i-j}
\right)-h\left(1-\frac {h}{2}\right)\cH_{i,j}-
\frac {\cJ_{i,j+1}-\cJ_{i,j-1}}{2}+\frac {\cH_{i,j+1}+\cH_{i,j-1}}{2}. &
\eeq
The boundary conditions  (\ref{eq:condbond}) imply that
\be \label{eq:bouncondition}
\cJ_{i,i}=\cH_{i,i}=0
\ee
for all $i$. The values of the unknown functions below
the boundary $t=\tau$ (at the points with $j=i-1$) indicated by asterisks in
Fig. 4, were obtained by interpolation using Newton's formula with three points
$j=i+1,i-1,i-2$:
\be \label{eq:cJnp1n}
\cJ_{i+1,i}=\cJ_{i+1,i-1}-\cJ_{i+1,i-2}/3,\quad
\cH_{i+1,i}=\cH_{i+1,i-1}-\cH_{i+1,i-2}/3.
\ee

The flux at the source is equal to zero, and the value of $\cJ$ at $\tau=0$ was
found by extrapolation. Thus, for all $i$, we have
\be \label{eq:cJn0}
\cH_{i,0}=\cH_{i+1,0}=0,\quad
\cJ_{i,0}=3(\cJ_{i,1}-\cJ_{i,2})+\cJ_{i,3}.
\ee

The following order of calculations was employed: first, for the chosen step
size $h$ the values of the unknown functions were calculated from their
expansions for $i=i_*$. Then a transition to larger values of $i$ was made
in succession: for $j=1, ...,i-1$ from $i$ to $i+1$, using Eqs.
(\ref{eq:Jip1})--(\ref{eq:Hip1}), for $j=i+1$ using Eqs.
(\ref{eq:bouncondition}), and for $j=0$ and $j=i$ using the values already
calculated according to Eqs. (\ref{eq:cJnp1n}) and (\ref{eq:cJn0}).

During the calculations, with increasing $i$ it was necessary to increase the
step size, taking the calculated values for the last $i$ as the initial values.
The former were taken to be $h=1/2^m$ , $t_*=1/8$($i_*=2^{m-3}$), with $m$
chosen to be $m=8-10$, depending on $t_{\s}$. At $t=5$ and then at $t=12$, the
step size was doubled. The calculation was continued up to some number $i=i_0$
when the estimated functions attained their asymptotes with the same accuracy
as the exact functions.

A comparison of the numerically determined values of $\cJ(t_i,\tau_j)$ and
$\cH(t_i,\tau_j)$ with those calculated with the exact formulas showed that for
$m=8$ and $t_{\s}=1$, the maximum relative error was $10^{-3}$, while for
$t_{\s}=2$ it was $10^{-4}$. The error increases as the parameter $t_{\s}$ is
reduced, since the function (\ref{eq:gtcont}) becomes narrower and higher
(closer to a $\delta$-function). When the allowed accuracy is reached, it is
necessary to reduce the step size.

A similar numerical scheme, based on a predictor-correction method, has been
developed by MacCormac \cite{Rouch}. It can also be used to solve these
problems.

\medskip

12. {\it Conclusion.} The exact solution of the equations obtained here
have made it possible to choose a method that is suitable for solving them
numerically and can be used to estimate the parameters required to achieve a
given accuracy. It is proposed that this method be used for calculating the
evolution of the luminosity and polarization of sources in the universe during
epochs close to the recombination epoch.

This work was partially supported by grant No. NSh-8542.2006.2 from the
President of the Russian Federation for leading scientific schools.

\end {document}